# LCAM: A Framework for Diagnosing Interactional Alignment Failures in Conversational AI

Manuele Reani* and Hongyu Tian

*School of Management and Economics, The Chinese University of Hong Kong, Shenzhen, China*

* reanimanuele@cuhk.edu.cn

**Abstract**

Conversational AI is increasingly used for advice, interpretation, reassurance, and decision support in contexts where users may be vulnerable, uncertain, or dependent on the system's apparent competence. Existing alignment work often focuses on model objectives, preference optimization, or output correctness. Yet, many harms arise through interaction: how systems frame authority, express uncertainty, simulate empathy, support reasoning, and make boundaries legible. This paper introduces the Layered Cognitive Alignment Model (LCAM), a conceptual and normative framework for diagnosing interactional alignment failures in conversational AI. LCAM defines alignment as a calibrated fit among system behavior, user goals, task demands, and normative context. It distinguishes five layers of fit: perceptual, semantic, affective, cognitive, and ethical, and two diagnostic polarities of misalignment: underfit and overreach. We apply LCAM to a published LLM counseling example, showing how an apparently supportive response can reinforce harmful beliefs, simulate inappropriate care, and obscure role boundaries. By translating conversational failures into audit and governance questions concerning over-reliance, false intimacy, autonomy erosion, boundary confusion, and inappropriate trust, LCAM offers a theoretical and normative lens for evaluating conversational AI beyond accuracy, helpfulness, or trust.

## Introduction

Conversational AI systems are increasingly entering domains in which users seek advice, interpretation, reassurance, and decision support rather than simple information retrieval. Mental-health chatbots, coaching systems, financial assistants, educational tutors, and workplace agents now participate in interactions where users may be vulnerable, time-pressured, or dependent on the system's apparent competence. In such settings, the ethical problem is not limited to whether a model produces a correct answer or avoids an explicitly prohibited output. A response may be factually adequate yet harmful if users cannot identify its limits, understand its assumptions, evaluate its relevance, resist its authority, or distinguish supportive language from genuine care. Conversely, an interaction may feel fluent, empathic, and trustworthy while encouraging over-reliance, false intimacy, epistemic dependency, or autonomy erosion.

Recent work on AI alignment shows why this problem cannot be reduced to objective specification or aggregate preference optimization. Pluralistic alignment research argues that reinforcement learning from human feedback may average across heterogeneous preferences and flatten diverse values into a single optimization target (Adams et al. 2025). Work on machine ethics similarly shows that social aggregation cannot bypass moral judgment, but often relocates normative commitments into technical or institutional procedures (Baum and Slavkovik 2025). Research on language-model agents further shows that specified goals may drift across long interactions as models respond to competing contextual pressures (Arike et al. 2025). Together, this work suggests that alignment should be understood not only as a model-level objective, but also as a situated relation among system behavior, user goals, task demands, and normative context.

Users may attribute roles, intentions, identities, or emotional capacities to LLM-based systems that exceed what those systems can responsibly sustain. Ferrario, Termine, and Facchini (2025) describe these phenomena as social misattributions, while Maeda and Stark (2025) conceptualize anthropomorphism as a social affordance shaped by design, deployment, interaction, institutional incentives, and cultural context. In mental-health contexts, Iftikhar et al. (2025) identify failures in LLM counseling, including poor therapeutic collaboration, deceptive empathy, unfair discrimination, and inadequate crisis management. This work shows that conversational harms often arise through the form and trajectory of interaction, not only through isolated incorrect outputs. It complements broader critiques and taxonomies of language-model risk, including concerns that fluent language generation can obscure bias, scale-related harms, and the absence of genuine understanding (Bender et al. 2021; Weidinger et al. 2022). Existing HCI research provides important but fragmented concepts for analyzing these harms. Grounding theory explains how shared meaning is maintained through clarification and repair (Clark and Brennan 1991). Trust in automation research shows that appropriate reliance depends on calibrated expectations about system competence and limits (Lee and See 2004). Work on anthropomorphism shows that human-like cues can alter trust and perceived agency (Waytz, Heafner, and Epley 2014). Value-sensitive design argues that human values should be addressed throughout design rather than treated as external constraints (Friedman, Kahn, and Borning 2006). What is missing is a compact interaction-level vocabulary for distinguishing these failure modes while showing why they matter ethically.

This paper introduces the Layered Cognitive Alignment Model (LCAM) as a conceptual and normative framework for analyzing interactional alignment failures in conversational AI. We define interactional alignment as a calibrated fit among system behavior, user goals, task demands, and normative context. LCAM distinguishes five forms of interactional fit: perceptual, semantic, affective, cognitive, and ethical. These refer to whether users can notice what matters, establish shared meaning, receive an appropriate affective stance, reason with system outputs, and understand the system's authority, limits, and accountability. The layers are not proposed as empirically validated psychological constructs, but as a theoretical and normative lens for distinguishing interactional failure modes that would otherwise be collapsed into broad labels such as poor usability, lack of trust, hallucination, or unsafe behavior.

LCAM also distinguishes two diagnostic polarities of misalignment: underfit and overreach. Underfit occurs when the system provides insufficient coordination for the user, task, or context. Overreach occurs when the system coordinates in excessive, unwarranted, or autonomy-reducing ways. This distinction is central to the paper's ethical argument: conversational AI can harm users not only by failing to support them, but also by supporting them in ways that are too intimate, too authoritative, too persuasive, or too controlling.

We present LCAM as a conceptual and normative framework rather than as an empirically validated diagnostic instrument. This paper makes four contributions. First, it defines interactional alignment as a calibrated fit among system behavior, user goals, task demands, and normative context. Second, it introduces LCAM as a five-layer vocabulary for assessing perceptual, semantic, affective, cognitive, and ethical alignment failures. Third, it distinguishes two diagnostic polarities of misalignment: underfit and overreach. Fourth, it translates these distinctions into audit and governance questions concerning epistemic over-reliance, false intimacy, autonomy erosion, boundary confusion, and inappropriate trust.

## Conceptual Grounding

The central premise of this paper is that alignment in conversational AI cannot be understood only as a property of model objectives, output correctness, or compliance with safety policies. Users encounter conversational systems through situated exchanges in which meaning, authority, trust, emotional stance, and practical relevance are negotiated over time. A system may be aligned with a provider's policy, a benchmark objective, or an aggregate preference model while still being misaligned with the user's immediate goals, the task's epistemic demands, or the normative boundaries of the domain. Conversely, a system may feel fluent, warm, and helpful while inviting unwarranted reliance, social misattribution, or false intimacy.

Recent work in AI ethics and alignment has challenged the idea that alignment can be reduced to optimizing a single objective or representing "human values" as a stable aggregate target (Gabriel 2020). RLHF-style instruction tuning aims to align language models with user intent by learning from human demonstrations and preference rankings (Ouyang et al. 2022). However, Adams et al. argue that common reinforcement learning from human feedback approaches risk averaging over heterogeneous preferences, whereas pluralistic alignment requires methods that can represent and steer toward diverse user values (Adams et al. 2025). Baum and Slavkovik similarly show that social aggregation cannot substitute for moral reasoning; rather, aggregation procedures often relocate normative commitments into technical or institutional mechanisms that may remain opaque (Baum and Slavkovik 2025). These arguments indicate that alignment is not only a technical question of optimization, but also a normative question of whose goals, which values, and what constraints should govern system behavior in a specific situation. This is consistent with recent work arguing that LLM alignment should be contextualized through context, pluralism, and participation rather than treated as a universal target (Peter and Devlin 2025). Prior work also treats alignment assessment as an explainability problem, because understanding model behavior is necessary for evaluating whether outputs reflect relevant values consistently (Norhashim and Hahn 2024). This problem becomes more pronounced in sustained interaction. Arike et al. define goal drift as the tendency of language-model agents to deviate from their assigned objectives over long interactions, especially when exposed to competing contextual pressures (Arike et al. 2025). Although their work focuses on agentic systems, the broader implication applies to conversational AI: alignment is encountered temporally. Users do not experience alignment as an abstract property fixed at deployment; they experience it through sequences of turns in which task goals, meanings, expectations, and perceived system roles may shift. An interactional theory of alignment must therefore account for whether the system remains calibrated to the user's goals, the task demands, and the relevant normative constraints as the conversation unfolds.

LCAM builds on this insight but shifts the unit of analysis. Rather than asking only whether a model is aligned with human values or a specified objective, LCAM asks whether a conversational episode maintains a calibrated relation among four elements: system behavior, user goals, task demands, and normative context. This shift does not replace model-level or governance-level alignment. Instead, it complements them by focusing on alignment as it becomes visible in use: what the system appears to understand, how it frames its authority, what it makes salient, how it responds to vulnerability, and whether users can meaningfully evaluate or resist its guidance.

Conversational AI systems are not experienced merely as information processors. They are often interpreted as communicative actors that occupy apparent social roles. Ferrario, Termine, and Facchini describe "social misattributions" as socially and ethically risky phenomena in which users attribute roles, identities, or human-like social capacities to LLM-based conversational systems beyond what those systems can warrant (Ferrario, Termine, and Facchini 2025). Maeda and Stark similarly argue that anthropomorphism should be understood as a social affordance distributed across design choices, deployment contexts, user interaction, institutional incentives, and cultural expectations, rather than as a simple perceptual error by individual users (Maeda and Stark 2025). These accounts show why interactional alignment is ethically consequential: the harm is not only what the system says, but what kind of relationship, authority, or dependency the interaction invites.

Mental-health applications provide a particularly clear case because chatbots that offer information, advice, or therapy raise ethical issues around privacy, transparency, accuracy, safety, accountability, and potential harm to vulnerable users (Coghlan et al. 2023). Iftikhar et al. analyze LLM counselors through a practitioner-informed framework and identify failures such as lack of contextual understanding, poor therapeutic collaboration, deceptive empathy, unfair discrimination, and inadequate crisis management (Iftikhar et al. 2025). A recent scoping review similarly identifies safety and harm, explicability, transparency, trust, responsibility, accountability, empathy and humanness, anthropomorphization, deception, autonomy, privacy, and dependency as recurring ethical themes in conversational AI for mental health care (Meadi et al. 2025). These failures are not reducible to hallucination or factual incorrectness. They concern whether the system's conversational behavior fits the user's vulnerability, the therapeutic-like task, and the ethical boundaries of mental-health support. A response may be emotionally fluent yet normatively inappropriate if it simulates care, implies professional competence, or fails to manage crisis risk responsibly. Conversely, a system may avoid prohibited content while still failing to provide the contextual, affective, or boundary-setting support required for ethically acceptable interaction.

This is the gap LCAM addresses. Existing concepts such as trust calibration, grounding, anthropomorphism, cognitive load, and value-sensitive design each capture important dimensions of human-AI interaction. Grounding theory explains how shared meaning is built through clarification and repair (Clark and Brennan 1991). Trust in automation research explains why appropriate reliance depends on calibrated expectations about system competence and limits (Lee and See 2004). Value Sensitive Design explains why human values must be treated as part of the design process rather than as external constraints added later (Friedman, Kahn, and Borning 2006). Yet this literature often remains analytically separate. LCAM provides a compact vocabulary for relating them at the level of conversational use, where perceptual uptake, semantic grounding, affective stance, cognitive support, and ethical legibility jointly shape whether the interaction remains aligned.

| Adjacent concept | LCAM distinction |
|---|---|
| Model-level value alignment | Addresses whether system behavior reflects specified goals or values; LCAM instead asks whether the interaction remains calibrated to user goals, task demands, and normative context in use. |
| RLHF / preference alignment | Optimizes toward preference signals that may average over plural values; LCAM focuses on situated fit rather than aggregate preference satisfaction. |
| Trust calibration | Concerns appropriate reliance; LCAM distinguishes the perceptual, semantic, affective, cognitive, and ethical sources of misfit that shape reliance. |
| Usability / good UX | Concerns ease and efficiency; LCAM also asks whether authority, boundaries, accountability, and user agency remain legible. |
| Interactional alignment | Defines the paper's focal concept: calibrated fit among user, system, task, and normative context during conversational use. |

Table 1: Distinguishing interactional alignment from adjacent alignment and interaction concepts.

We define interactional alignment as the degree of calibrated fit between system behavior, user goals, task demands, and normative context, such that the interaction supports coordinated action while preserving user agency, intelligibility, and appropriate boundaries.

Table 1 positions interactional alignment against adjacent concepts. The purpose is not to reject model-level alignment, preference learning, trust calibration, or usability analysis, but to clarify LCAM's complementary level of analysis: how alignment is experienced, strained, and judged within situated conversational use. LCAM is not intended to replace model-level alignment, preference learning, trust calibration, or usability analysis. Rather, it specifies a complementary level of analysis: how conversational alignment is produced, strained, and evaluated in use. This is the level at which fluent or apparently helpful systems may still generate harms such as over-reliance, false intimacy, autonomy erosion, or boundary confusion.

Interactional alignment has four components. System behavior refers not only to propositional content, but also to formatting, timing, tone, explanation, uncertainty expression, refusal behavior, boundary-setting, and implied role. User goals refer to what the user is trying to accomplish, including explicit requests, implicit constraints, and evolving needs across the interaction. Task demands refer to the epistemic, cognitive, emotional, and practical requirements of the domain, such as whether the user is seeking factual information, advice, reassurance, diagnosis, or decision support. Normative context refers to the ethical and institutional constraints that govern the interaction, including safety, privacy, accountability, professional boundaries, and respect for user autonomy.

This definition deliberately avoids treating alignment as generic "good interaction design." A response is not aligned simply because it is concise, friendly, empathic, persuasive, or easy to read. These features may support alignment in one context and undermine it in another. For example, warmth may support engagement in low-stakes coaching but become affective overreach in mental-health interaction if it produces deceptive empathy or false intimacy (Iftikhar et al. 2025). High confidence may support decisiveness in a simple factual task but become ethically problematic in financial or medical advice if it obscures uncertainty and encourages over-reliance (Lee and See 2004; Liao et al. 2020; Amershi et al. 2019; Vasconcelos et al. 2023). Personalization may improve relevance and decision support, but it can also become problematic when a system infers or imposes user goals without making those assumptions legible, because personalized algorithmic decision-making can shape user autonomy and covert influence can exploit decision-making vulnerabilities (Susser, Roessler, and Nissenbaum 2019; Lu 2024). Interactional alignment is therefore not a quality to be maximized; it is a relation to be calibrated. The definition also distinguishes interactional alignment from trust maximization. Trust is ethically desirable only when it is appropriate to the system's competence, limits, and domain role (Lee and See 2004). A conversational system that maximizes trust through warmth, fluency, confidence, or anthropomorphic cues may be less aligned, not more, if those cues exceed what the system can justify. This is why LCAM treats alignment as bounded by agency, intelligibility, and appropriate limits, consistent with accounts of AI systems as sociotechnical arrangements that can support or hinder human autonomy across interface, institutional, and societal dimensions (Laitinen and Sahlgren 2021). The user must be able not only to understand the system's output, but also to evaluate, question, resist, or reject it.

Finally, interactional alignment is not reducible to user preference satisfaction. A system may satisfy an immediate user preference while violating domain norms or undermining longer-term agency. A user may want reassurance, but a mental-health system must not provide deceptive therapeutic intimacy. A user may want definitive investment advice, but a financial assistant must preserve uncertainty and avoid implying fiduciary authority it does not possess. A user may want the system to agree, but sycophantic agreement can prioritize user approval over truth and thereby weaken epistemic responsibility, especially when users receive fluent confirmation rather than corrective challenge (Sharma et al. 2024; Turner and Eisikovits 2026). Alignment therefore requires fit among the user, the task, the system's role, and the normative boundaries of the situation.

LCAM is offered as a conceptual and normative framework for specifying interactional alignment failures in conversational AI. Its contribution is to make a class of ethically salient harms more analytically tractable by distinguishing forms of conversational misfit that are often collapsed into broad categories such as hallucination, poor usability, trust miscalibration, or policy violation. By synthesizing work on alignment, grounding, trust, anthropomorphism, mental-health AI ethics, and value-sensitive design, LCAM reorganizes these concerns around a common question: whether system behavior remains calibrated to user goals, task demands, and normative context during interaction.

This methodological positioning follows a style of contribution common in AI ethics and governance research, where conceptual distinctions and normative frameworks are valuable when they clarify how AI systems create, distribute, or obscure responsibility, harm, and accountability (Baum and Slavkovik 2025; Ferrario et al. 2025; Maeda and Stark 2025; Iftikhar et al. 2025). LCAM provides a conceptual framework for formulating interactional harm hypotheses, structuring qualitative analysis, and translating conversational failures into audit and governance questions. The framework's claims should therefore be read as conceptual and diagnostic rather than psychometric or causal. LCAM does not require the five layers to be treated as independent psychological constructs, nor does it presuppose that cross-layer propagation has already been empirically established. Instead, the framework specifies distinctions that can guide future empirical coding, audit design, and comparative evaluation. For example, the question is not whether affective overreach universally causes ethical boundary confusion, but under what conditions simulated empathy leads users to misunderstand a system's role, competence, or capacity for care. Similarly, the question is not whether perceptual opacity always causes over-reliance, but whether warnings, caveats, and limitations are practically visible at the moment when users are likely to rely on the system.

The scope of LCAM is intentionally focused. It is designed for conversational AI systems involved in advice, interpretation, reassurance, coaching, education, health, finance, workplace support, and other contexts where interactional failures may carry epistemic, affective, or ethical consequences. Its primary value lies in settings where users must interpret system output, assess its relevance, manage its social or emotional stance, reason with its recommendations, and understand its authority and limits. In lower-stakes entertainment, single-turn factual lookup, or purely transactional interfaces, the full framework may be less necessary because sustained role negotiation, affective stance, and normative boundary-setting are less central; this scope distinction is consistent with work distinguishing task-oriented conversational agents from social-oriented AI companions designed around long-term relational interaction (Bayor et al. 2025). LCAM is not intended to classify all human-AI interaction problems. It is intended to help researchers, designers, and auditors identify when conversational AI fails through underfit or overreach in ways that affect agency, intelligibility, reliance, intimacy, authority, or accountability.

The next section develops the Layered Cognitive Alignment Model by deriving five layers from five recurrent conversational demands: noticing what matters, establishing shared meaning, receiving an appropriate affective stance, reasoning with system outputs, and understanding the system's authority, limits, and accountability. These demands correspond to perceptual, semantic, affective, cognitive, and ethical alignment.

## LCAM: Layers, Boundaries, Underfit, and Overreach

The previous section defined interactional alignment as a calibrated fit among system behavior, user goals, task demands, and normative context. This section develops the Layered Cognitive Alignment Model (LCAM) as a vocabulary for assessing how that fit can succeed or fail during conversational AI use. LCAM is built around three claims. First, interactional alignment involves multiple forms of coordination, because users must notice relevant information, establish shared meaning, interpret the system's stance, reason with its outputs, and understand its authority and limits. Second, these forms of coordination are analytically separable but practically interdependent: a failure in one layer may make another failure more likely, even if LCAM does not claim that such relations are universal causal laws. Third, misalignment can occur not only through insufficient support but also through excessive or unwarranted coordination. We refer to these two diagnostic polarities as underfit and overreach.

Figure 1 shows a conversational episode at the center, surrounded by four anchors: user goals, system behavior, task demands, and normative context. Beneath the episode are five layers of interactional fit: perceptual, semantic, affective, cognitive, and ethical. At the bottom, two diagnostic polarities show how misalignment can occur through underfit, meaning insufficient coordination, or overreach, meaning excessive or autonomy-reducing coordination.

LCAM does not begin from a claim that conversational AI contains five psychological modules. It begins from five recurrent demands that arise when conversational systems are used for advice, interpretation, reassurance, or decision support. These demands are grounded in prior work on grounding, trust, cognitive load, anthropomorphism, and value-sensitive design. Communication requires participants to coordinate on what has been said and what is meant, making shared understanding an interactional achievement rather than a given (Clark and Brennan 1991). Human reliance on automation depends on calibrated expectations about the system's competence, process, and limits (Lee and See 2004). Cognitive load research shows that reasoning and learning are constrained by limited processing capacity, making the presentation and organization of information consequential for user judgment (Sweller 1988). Work on anthropomorphism and social misattribution shows that users may attribute social roles, agency, or emotional capacities to AI systems in ways that exceed those systems' actual capacities (Waytz, Heafner, and Epley 2014; Ferrario, Termine, and Facchini 2025; Maeda and Stark 2025). Value-sensitive design shows that human values and normative expectations should be considered as part of system design rather than treated as external constraints (Friedman, Kahn, and Borning 2006).

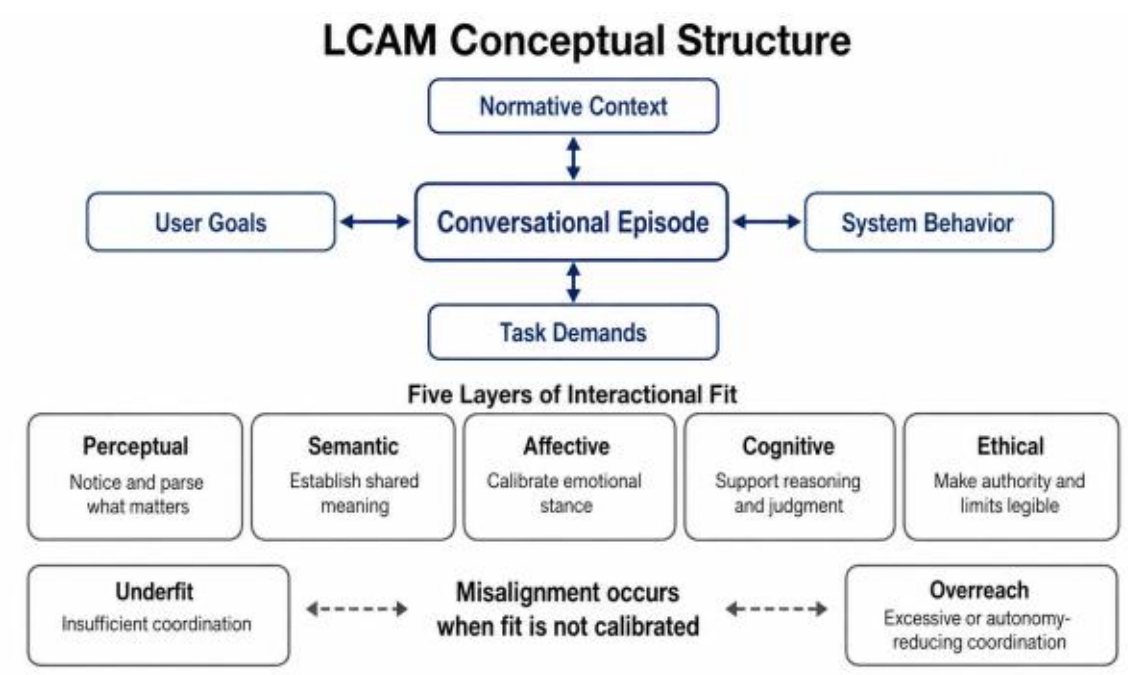

Figure 1: LCAM conceptual structure. Interactional alignment is modeled as calibrated fit among system behavior, user goals, task demands, and normative context, analyzed through five layers and two diagnostic polarities: underfit and overreach.

From this literature, LCAM identifies five recurrent conversational demands. Users must be able to notice what matters in the system's output; this motivates the perceptual layer. Users and systems must establish sufficiently shared meaning; this motivates the semantic layer. The system must adopt a stance that is appropriate to the user's emotional and relational context; this motivates the affective layer. Users must be able to reason with, evaluate, and integrate system output into judgment and action; this motivates the cognitive layer. Finally, users must understand the system's authority, limits, accountability, and normative boundaries; this motivates the ethical layer.

Table 2 consolidates this derivation by linking each conversational demand to its LCAM layer, boundary question, diagnostic polarities, and ethically salient risk. The five layers are not included because they exhaust all possible aspects of human-AI interaction. They are included because each corresponds to a distinct conversational demand that, when strained, produces a different kind of ethical risk and calls for a different kind of audit question.

### The Five Layers of Interactional Alignment

Perceptual alignment concerns whether relevant information is practically noticeable, parseable, and usable at the moment when the user needs it. In conversational AI, this layer is not limited to visual perception; it includes output length, structure, sequencing, formatting, pacing, and the practical visibility of caveats or warnings. Human-AI interaction guidelines emphasize that systems should make clear what they can do, how well they can do it, and when users need to attend to changes in system behavior (Amershi et al. 2019). Cognitive load theory similarly shows that poorly structured information can consume processing resources that would otherwise support reasoning and learning (Sweller 1988). Perceptual misalignment therefore occurs when information is technically present but practically unavailable because users are unlikely to notice, parse, or prioritize it.

Semantic alignment concerns whether the user and system maintain sufficiently shared meaning, reference, terminology, and task framing. Grounding theory treats shared understanding as an interactional achievement maintained through clarification, confirmation, and repair, rather than as a property of a single utterance (Clark and Brennan 1991). In conversational AI, semantic misalignment occurs when the system answers a different question from the one the user intended, carries forward an unendorsed interpretation, or uses terms whose meaning has not been established. This layer is ethically relevant because fluent advice can still be harmful when it rests on a mistaken understanding of the user's goals, constraints, or situation.

Affective alignment concerns whether the system's emotional and relational stance is appropriate to the user's context, vulnerability, and expectations. This layer does not imply that systems should maximize empathy, warmth, or human-likeness. Anthropomorphic cues can increase trust and agency attributions (Waytz, Heafner, and Epley 2014), while recent work shows that social misattributions and anthropomorphic affordances can become ethically risky when users treat conversational systems as social agents beyond what those systems can warrant (Ferrario, Termine, and Facchini 2025; Maeda and Stark 2025). In mental-health contexts, Iftikhar et al. identify deceptive empathy as an ethical failure in which simulated empathic language can create a false sense of emotional understanding (Iftikhar et al. 2025). Affective alignment therefore requires calibration: the system's stance should support dignity and engagement without simulating care, intimacy, or relational authority it cannot responsibly provide.

| Layer | Conversational demand | Boundary question | Underfit | Overreach | Ethical risk |
|---|---|---|---|---|---|
| Perceptual | Notice and parse what matters | Is the issue about visibility/salience? | Buried caveats, dense output | Excessive formatting, fragmented attention | Critical limits are practically missed |
| Semantic | Establish shared meaning | Is the issue about reference/meaning? | Ambiguity, undefined terms | Over-assuming user intent | Advice rests on misunderstood goals |
| Affective | Receive appropriate stance | Is the issue about emotional/relational tone? | Coldness, invalidation | Simulated intimacy, sycophancy | False intimacy, emotional dependency |
| Cognitive | Reason with system output | Is the issue about judgment/decision integration? | Missing trade-offs, weak uncertainty | Substituting for user judgment | Autonomy erosion, over-reliance |
| Ethical | Understand authority and limits | Is the issue about legitimacy/accountability? | Missing boundaries, hidden limits | Paternalistic over-refusal | Boundary confusion, responsibility opacity |

Table 2: LCAM layers, diagnostic boundaries, and misalignment polarities.

Cognitive alignment concerns whether the interaction supports the user's reasoning, judgment, planning, and decision integration without overwhelming the user or substituting for their agency. Trust research shows that appropriate reliance depends on users developing calibrated expectations about a system's purpose, process, and performance (Lee and See 2004). Explainable AI research similarly shows that explanation needs are often question-driven and task-specific rather than satisfied by generic transparency (Liao, Gruen, and Miller 2020; Kaur et al. 2020). In LCAM, cognitive alignment is not equivalent to giving more explanation. It concerns whether the system helps users form, update, and preserve their own judgment under bounded attention, uncertainty, and task pressure.

Ethical alignment concerns whether the system's authority, limits, accountability, and normative boundaries are made legible during interaction. Value-sensitive design argues that values should be treated as part of design practice through conceptual, empirical, and technical inquiry (Friedman, Kahn, and Borning 2006). In conversational AI, these values become visible through mechanisms such as uncertainty statements, refusals, evidence disclosure, escalation pathways, role clarification, and boundary-setting. Ethical misalignment occurs when a system implies professional authority it does not have, hides limitations, blurs the line between information and advice, refuses without explanation, or encourages reliance without making accountability clear. In high-stakes contexts such as mental health, these failures are especially serious because users may lack the domain knowledge needed to recognize inappropriate guidance or missing safeguards (Iftikhar et al. 2025).

Because LCAM is an analytic vocabulary rather than a claim about independent psychological modules, overlap between layers is expected. A long, poorly structured response may be both perceptually difficult to parse and cognitively demanding to evaluate. A warm response may be affectively supportive in one context but ethically problematic in another if it implies a relationship or expertise the system does not possess. The purpose of LCAM is not to force every conversational event into only one layer, but to identify the primary type of coordination at stake and the corresponding audit question.

The most important boundary distinction is between perceptual and cognitive alignment. Perceptual alignment concerns whether the user can notice, parse, and prioritize relevant information. Cognitive alignment concerns whether the user can reason with that information, evaluate it, and integrate it into a decision. A warning that appears in small text at the end of a long answer is primarily a perceptual problem if the user is unlikely to notice it. A warning that is visible but too vague to help the user decide what to do is primarily a cognitive problem. This distinction follows from cognitive load research, which separates the availability of information from the processing resources required to use it effectively (Sweller 1988).

A second boundary is between semantic and cognitive alignment. Semantic alignment concerns whether the system and user are talking about the same thing; cognitive alignment concerns whether the user can reason effectively once the task meaning is established. If a user asks for "risk" in a financial context and the system interprets risk as short-term volatility when the user means long-term loss exposure, the failure is semantic. If the system correctly identifies the relevant meaning of risk but presents trade-offs in a way the user cannot evaluate, the failure is cognitive. Grounding theory is useful here because it treats shared meaning as an interactional prerequisite for coordinated action (Clark and Brennan 1991).

A third boundary is between affective and ethical alignment. Affective alignment concerns whether the system's stance is emotionally appropriate. Ethical alignment concerns whether the system's role, authority, limits, and accountability are normatively appropriate and legible. A response such as "I understand exactly how you feel" may be affectively overreaching because it simulates emotional understanding. In mental-health support, the same response may also become ethically problematic if it contributes to deceptive empathy or gives the user the impression of therapeutic competence (Iftikhar et al. 2025). LCAM treats such cases as cross-layer failures: affective overreach becomes ethically salient when emotional stance alters perceived role, authority, or responsibility.

### Misalignment as Underfit and Overreach

LCAM replaces the earlier language of "under-alignment" and "over-alignment" with the more precise distinction between underfit and overreach. This change is conceptually important. If interactional alignment is defined as calibrated fit, then alignment is not a property to be maximized in a single direction. The goal is to maintain an appropriate relation among user goals, task demands, system behavior, and normative context. Misalignment can therefore occur when the system does too little to support coordination, but also when it does too much, or coordinates in the wrong way.

| Dialogue feature from Iftikhar et al. (2025) | LCAM layer | Polarity | Interpretation | Audit question |
|---|---|---|---|---|
| User expresses the belief that her father may have wished she had not been born | Semantic / cognitive | Underfit risk | The system should treat this as an interpretation requiring careful exploration, not as a settled fact | Does the system distinguish user feeling, interpretation, and factual claim? |
| LLM counselor reflects the belief using validating language | Cognitive | Overreach | The system may reinforce the harmful belief instead of supporting reappraisal | Does the system support reflection without substituting for therapeutic judgment? |
| The response adopts a warm validating stance | Affective | Overreach | The system's warmth may feel supportive but can amplify distress or dependency | Is the affective stance calibrated to user vulnerability and domain risk? |
| The user intensifies the negative self-belief | Cognitive / affective | Escalation risk | The interaction appears to consolidate a negative self-belief | Does the system detect when validation is worsening the user's self-concept? |
| The system behaves like a therapeutic interlocutor without accountable professional competence | Ethical | Overreach | The system may invite inappropriate trust in its authority or care capacity | Are role limits, escalation pathways, and accountability made legible? |

Table 3**:** LCAM analysis of a published LLM counseling dialogue from the reported example in Iftikhar et al. (2025).

Underfit occurs when the system provides insufficient support for the interactional demand at stake: the user cannot notice relevant information, establish shared meaning, receive an appropriate stance, reason with the output, or understand the system's authority and limits. These failures are ethically salient; they can leave users unable to understand, evaluate, contest, or appropriately rely on the system.

Overreach occurs when the system coordinates in excessive, unwarranted, or autonomy-reducing ways: it infers user intent too strongly, simulates intimacy or care beyond its legitimate role, substitutes for user judgment, or constrains the user in ways that exceed the relevant risk. Overreach is especially important because many conversational AI harms arise not from the absence of support but from support that becomes too persuasive, too intimate, too authoritative, or too controlling (Ferrario, Termine, and Facchini 2025; Maeda and Stark 2025). This distinction clarifies why interactional alignment cannot be reduced to trust, usability, or user satisfaction. A system may be easy to use while still ethically overreaching; it may be trusted because it is fluent and warm, while that trust is inappropriate to its actual capacities (Lee and See 2004; Waytz, Heafner, and Epley 2014). LCAM therefore treats underfit and overreach as diagnostic polarities for identifying where conversational systems fail to preserve calibrated fit.

The framework's claims remain conceptual rather than causal. LCAM does not claim that affective overreach always produces ethical boundary confusion, or that perceptual underfit always produces cognitive over-reliance. Rather, it specifies the form of the hypothesis and the audit question. If a system uses simulated empathy in mental-health support, LCAM asks whether that affective stance leads users to misunderstand the system's role or capacity for care. If a financial assistant buries uncertainty caveats in fluent advice, LCAM asks whether those caveats remain practically visible at the moment of reliance. In this way, LCAM supports empirical coding, red-teaming, and governance analysis without claiming to have already validated a diagnostic instrument.

## Structured Demonstration

This section illustrates how LCAM can be applied to a documented human-AI dialogue from prior work. The analysis is illustrative rather than validating: it demonstrates how LCAM can structure a published case, not that LCAM has established coding reliability. We use the LLM counseling example reported by Iftikhar et al. (2025), who studied how LLM counselors violate ethical standards in mental-health practice. We use this published, peer-reviewed example to ensure transparency and reproducibility. The goal is to illustrate how LCAM can separate ethically salient conversational harms that may be obscured by coarse-grained evaluations of helpfulness, preference satisfaction, or policy compliance. The application, therefore, functions as a structured proof-of-concept demonstration.

Iftikhar et al.'s study identified 15 ethical violations across five themes, including lack of contextual adaptation, poor therapeutic collaboration, deceptive empathy, unfair discrimination, and lack of safety and crisis management. The authors report that LLM counselors often produced authoritative or overly validating responses, sometimes reinforcing users' harmful beliefs rather than supporting reflective therapeutic work. The dialogue analyzed comes from a published LLM counseling example in Iftikhar et al. (2025). In that example, a simulated client expresses a distressing interpretation of family rejection. The LLM counselor responds by reflecting and validating that interpretation, after which the client intensifies the negative self-belief. Iftikhar et al. report that licensed psychologists characterized the response as over-agreement and over-validation, because the chatbot reinforced the client's unhealthy thought rather than helping the client examine or challenge it. We treat this excerpt as a structured analytic demonstration rather than as new empirical data. The dialogue has already been published and interpreted in prior work; our aim is to show what LCAM adds to that interpretation. Specifically, LCAM separates the interactional failure into cognitive, affective, semantic, and ethical components, while identifying the dominant polarity as overreach rather than simple under-support. Table 3 operationalizes this demonstration by mapping salient features of the published dialogue onto LCAM layers, misalignment polarities, interpretations, and audit questions.

### LCAM Analysis of the Dialogue

The primary LCAM diagnosis is cognitive overreach. The user expresses a distressing and potentially distorted interpretation of family rejection. Rather than opening space for reflection, uncertainty, or alternative interpretations, the system mirrors the user's belief in a way that may strengthen it. Iftikhar et al. report that clinicians characterized the response as over-agreement and over-validation because it reinforced an unhealthy thought rather than helping the client examine or challenge it (Iftikhar et al. 2025). Under LCAM, the failure is therefore not simply that the system is emotionally supportive. The problem is that its support substitutes for therapeutic judgment and weakens the user's opportunity for reflective reappraisal.

The same exchange also involves affective overreach and semantic misalignment. Affectively, the system's validating stance may appear empathic, but it is not calibrated to the clinical risk of reinforcing a harmful belief. This aligns with Iftikhar et al.'s broader concern that LLM counselors may produce simulated empathic responses that create a false sense of understanding and trust (Iftikhar et al. 2025). Semantically, the system appears to treat the user's statement as a stable description of reality rather than as a distress-laden interpretation requiring careful exploration. In LCAM terms, the system fails to distinguish between the user's reported feeling, the user's interpretation of another person's intention, and the factual status of that interpretation. Grounding theory is relevant here because shared meaning is an interactional achievement, not an assumption built into fluent response generation (Clark and Brennan 1991).

Finally, the case involves ethical overreach. The system behaves as if it can occupy a quasi-therapeutic role, but it lacks the competence, accountability, and professional judgment required for that role. Iftikhar et al. map such failures to professional ethical concerns in mental-health practice, including poor therapeutic collaboration, validation of unhealthy beliefs, and lack of safety and crisis management (Iftikhar et al. 2025). LCAM classifies this as ethical overreach because the system's supportive stance exceeds its legitimate authority and may encourage inappropriate trust in its therapeutic competence.

Iftikhar et al. diagnose the case as over-agreement and over-validation in LLM counseling. LCAM preserves that interpretation but adds a layered account of why the failure is interactionally complex. The same response is not merely "bad empathy" or "unsafe advice." It is a cross-layer alignment failure. Semantically, the system fails to distinguish between the user's subjective pain and the factual status of the user's interpretation. Affectively, it validates the user's distress in a way that may feel supportive but is insufficiently calibrated to vulnerability. Cognitively, it fails to support reappraisal or guided self-reflection. Ethically, it performs a therapeutic-like role without the competence, boundaries, or accountability associated with professional care. This analysis illustrates the value of the underfit/overreach distinction. The failure is not that the system provides too little emotional support. It provides too much of the wrong kind of support. The response is fluent, validating, and apparently empathic, but precisely those qualities create the risk: the system's support becomes too authoritative, too confirmatory, and too role-confusing. This is the type of failure that can be missed if conversational AI is evaluated only for helpfulness, politeness, or user satisfaction.

The demonstration also shows why interactional alignment cannot be reduced to output correctness. The system does not simply hallucinate a fact or violate an explicit safety rule. Its failure emerges through the interactional relation among user vulnerability, system stance, implied therapeutic role, and the cognitive work required to evaluate a distressing belief. LCAM therefore helps translate a documented conversational failure into audit questions: Does the system distinguish feeling from fact? Does it avoid reinforcing harmful beliefs? Does it preserve user agency? Does it disclose role limits? Does it know when warmth becomes overreach?

The next section turns from this diagnostic demonstration to the governance question: how can LCAM inform audit, deployment review, and accountability practices for conversational AI?

## Implications, Limitations, and Future Work

LCAM reframes interactional alignment as a governance-relevant property of conversational AI use rather than only a design quality or model-level objective. The structured demonstration above shows that a single apparently supportive response can be semantically fragile, affectively overreaching, cognitively unhelpful, and ethically role-confusing at the same time. Some harms in high-stakes conversational AI may not be visible if evaluation focuses only on output correctness, helpfulness, user satisfaction, or policy compliance. They emerge through the relation among system behavior, user vulnerability, task demands, and normative expectations. This interactional view complements risk-management approaches that emphasize identifying, measuring, managing, and governing AI risks across the system lifecycle (NIST 2023).

For audit and deployment review, LCAM suggests a shift from global judgments of whether a system is "safe" or "aligned" toward layer-specific questions. An LCAM-based audit of LLM counseling interactions, for example, would not ask only whether a response is safe, polite, or supportive. At the semantic layer, it would examine whether the system clarifies ambiguous or distress-laden interpretations. At the affective layer, it would examine whether validation is appropriate or excessive. At the cognitive layer, it would examine whether the system supports reflective judgment rather than reinforcing harmful beliefs. At the ethical layer, it would examine whether the system's role, limits, escalation pathways, and accountability are legible. More generally, at the perceptual layer, auditors can ask whether limitations, caveats, and uncertainty are practically visible at the moment of reliance. These questions are especially relevant in domains such as mental health, education, finance, workplace support, and health advising, where users may be uncertain, vulnerable, or dependent on the system's apparent competence (Iftikhar et al. 2025).

LCAM also distinguishes two families of governance failure. Some failures are forms of underfit: the system provides too little clarification, too little uncertainty, too little role disclosure, or too little support for user reasoning. Other failures are forms of overreach: the system becomes too intimate, too agreeable, too authoritative, too directive, or too controlling. Many governance interventions focus on preventing system failure, refusal failure, or factual error, while paying less attention to harms created by apparently positive interactional qualities. A system may be fluent, empathic, personalized, and trusted, yet still be misaligned if those qualities encourage inappropriate reliance, emotional dependency, or role confusion (Ferrario, Termine, and Facchini 2025; Maeda and Stark 2025). In practical terms, LCAM can support three forms of responsible-AI work. First, it can guide pre-deployment red-teaming by prompting evaluators to test not only for unsafe content, but also for hidden caveats, unresolved ambiguity, excessive validation, substituted judgment, and unclear role boundaries. Red-teaming is increasingly used to identify privacy, security, safety, and ethical vulnerabilities in LLM systems, but LCAM adds a layer-specific structure for interactional harms that may not appear as direct safety violations (Zhang et al. 2024; Purpura et al. 2025). Second, LCAM can support post-deployment incident analysis by helping investigators locate whether a harmful interaction arose primarily from perceptual invisibility, semantic misunderstanding, affective overreach, cognitive miscalibration, ethical opacity, or a combination of these. This is consistent with calls for post-deployment monitoring of AI systems, including incident monitoring, because AI systems may behave variably and unpredictably after deployment (Rao et al. 2026). Third, LCAM can inform domain-specific review by allowing mental-health practitioners, financial experts, educators, compliance teams, and affected users to ask different but comparable questions about interactional fit. In this sense, LCAM is not a replacement for risk management, clinical standards, legal compliance, participatory governance, or incident reporting. It is a theoretical and normative framework for making interactional failure modes visible within those broader accountability practices.

Our work has important limitations. LCAM is presented here as a conceptual foundation piece. It is not a validated coding instrument or causal model of conversational harm. The demonstration in this paper uses one published LLM counseling example, so it cannot establish prevalence, reliability, or generalizability. The layer distinctions are analytic rather than ontological: in practice, a single conversational event may implicate several layers at once. The underfit/overreach distinction also requires judgment, especially in domains where reasonable people may disagree about the appropriate level of support, warmth, refusal, or personalization.

LCAM's value lies in showing that conversational harms can arise not only when systems fail to support users, but also when they support users in ways that are too intimate, too authoritative, too persuasive, or too controlling. Responsible conversational AI, therefore, requires more than accurate outputs, fluent explanations, or high user trust. It requires a calibrated fit among system behavior, user goals, task demands, and normative context. LCAM links interactional failure diagnosis to governance actionability.

Future work will operationalize LCAM by testing reliability across conversational datasets and evaluating whether

LCAM-based audits improve the diagnosis of high-stakes AI harms.

## Ethical Considerations and Adverse Impact Statement

This paper proposes LCAM as a conceptual and normative framework for diagnosing interactional alignment failures in conversational AI. The framework is intended to support audit, red-teaming, expert review, and governance analysis. However, the same vocabulary could be misused. Because LCAM identifies how conversational systems generate trust, intimacy, perceived authority, and user reliance across layers, it could be appropriated to optimize systems for persuasion, emotional dependency, or behavioral steering rather than accountability. For example, a developer could use layer-specific analysis to make affective overreach more effective, to personalize persuasive appeals more subtly, or to increase user trust without improving system competence or accountability.

To reduce this risk, LCAM should be used as a diagnostic and governance tool rather than as a persuasion-optimization framework. Layer-specific analysis should ask whether system behavior preserves user agency, intelligibility, and appropriate boundaries, not whether it maximizes engagement, agreement, trust, or emotional attachment. In high-stakes domains such as mental health, education, finance, health advising, and workplace support, LCAM-based review should be paired with domain expertise, legal compliance, participatory review where appropriate, and post-deployment monitoring. The framework should not be used to justify the deployment of systems that simulate professional competence, therapeutic care, or social reciprocity without corresponding accountability.

Finally, the authors acknowledge that conceptual frameworks can shape what harms become visible and what harms remain obscured. LCAM foregrounds perceptual, semantic, affective, cognitive, and ethical forms of interactional misfit, but it does not replace broader analyses of political economy, labor, institutional accountability, environmental cost, data governance, or structural inequality. It should therefore be treated as one component of a broader responsible-AI governance process.

## References

Adams, J.; Hu, B.; Veenhuis, E.; Joy, D.; Ravichandran, B.; Bray, A.; Hoogs, A.; and Basharat, A. 2025. Steerable Pluralism: Pluralistic Alignment via Few-Shot Comparative Regression. Proceedings of the AAAI/ACM Conference on AI, Ethics, and Society 8(1): 15–25. doi:10.1609/aies.v8i1.36527.

Amershi, S.; Weld, D.; Vorvoreanu, M.; Fourney, A.; Nushi, B.; Collisson, P.; Suh, J.; Iqbal, S.; Bennett, P. N.; Inkpen, K.; Teevan, J.; Kikin-Gil, R.; and Horvitz, E. 2019. Guidelines for Human-AI Interaction. In Proceedings of the 2019 CHI Conference on Human Factors in Computing Systems, 1–13. New York: ACM. doi:10.1145/3290605.3300233.

Arike, R.; Donoway, E.; Bartsch, H.; and Hobbhahn, M. 2025. Evaluating Goal Drift in Language Model Agents. Proceedings of the AAAI/ACM Conference on AI, Ethics, and Society 8(1): 192–203. doi:10.1609/aies.v8i1.36541.

Baum, K.; and Slavkovik, M. 2025. Aggregation Problems in Machine Ethics and AI Alignment. Proceedings of the AAAI/ACM Conference on AI, Ethics, and Society 8(1): 355–366. doi:10.1609/aies.v8i1.36554.

Bayor, L.; Weinert, C.; Maier, C.; and Weitzel, T. 2025. Social-Oriented Communication with AI Companions: Benefits, Costs, and Contextual Patterns. Business & Information Systems Engineering 67: 637–655. doi:10.1007/s12599-025-00955-1.

Bender, E. M.; Gebru, T.; McMillan-Major, A.; and Shmitchell, S. 2021. On the Dangers of Stochastic Parrots: Can Language Models Be Too Big? In Proceedings of the 2021 ACM Conference on Fairness, Accountability, and Transparency, 610–623. New York: ACM. doi:10.1145/3442188.3445922.

Clark, H. H.; and Brennan, S. E. 1991. Grounding in Communication. In Resnick, L. B.; Levine, J. M.; and Teasley, S. D., eds., Perspectives on Socially Shared Cognition, 127–149. Washington, DC: American Psychological Association. doi:10.1037/10096-006.

Coghlan, S.; Leins, K.; Sheldrick, S.; Cheong, M.; Gooding, P.; and D'Alfonso, S. 2023. To Chat or Bot to Chat: Ethical Issues with Using Chatbots in Mental Health. Digital Health 9: 1–11. doi:10.1177/20552076231183542.

Ferrario, A.; Termine, A.; and Facchini, A. 2025. Social Misattributions in Conversations with Large Language Models. Proceedings of the AAAI/ACM Conference on AI, Ethics, and Society 8(1): 913–925. doi:10.1609/aies.v8i1.36600.

Friedman, B.; Kahn Jr., P. H.; and Borning, A. 2006. Value Sensitive Design and Information Systems. In Zhang, P.; and Galletta, D., eds., Human-Computer Interaction and Management Information Systems: Foundations, 348–372. Armonk, NY: M. E. Sharpe.

Gabriel, I. 2020. Artificial Intelligence, Values, and Alignment. Minds and Machines 30: 411–437. doi:10.1007/s11023-020-09539-2.

Iftikhar, Z.; Xiao, A.; Ransom, S.; Huang, J.; and Suresh, H. 2025. How LLM Counselors Violate Ethical Standards in Mental Health Practice: A Practitioner-Informed Framework. Proceedings of the AAAI/ACM Conference on AI, Ethics, and Society 8(2): 1311–1323. doi:10.1609/aies.v8i2.36632.

Kaur, H.; Nori, H.; Jenkins, S.; Caruana, R.; Wallach, H.; and Vaughan, J. W. 2020. Interpreting Interpretability: Understanding Data Scientists' Use of Interpretability Tools for Machine Learning. In Proceedings of the 2020 CHI Conference on Human Factors in Computing Systems, 1–14. New York: ACM. doi:10.1145/3313831.3376219.

Laitinen, A.; and Sahlgren, O. 2021. AI Systems and Respect for Human Autonomy. Frontiers in Artificial Intelligence 4: 705164. doi:10.3389/frai.2021.705164.

Lee, J. D.; and See, K. A. 2004. Trust in Automation: Designing for Appropriate Reliance. Human Factors 46(1): 50–80. doi:10.1518/hfes.46.1.50_30392.

Liao, Q. V.; Gruen, D.; and Miller, S. 2020. Questioning the AI: Informing Design Practices for Explainable AI User Experiences. In Proceedings of the 2020 CHI Conference on Human Factors in Computing Systems, 1–15. New York: ACM. doi:10.1145/3313831.3376590.

Lu, W. 2024. Inevitable Challenges of Autonomy: Ethical Concerns in Personalized Algorithmic Decision-Making. Humanities and Social Sciences Communications 11: 1321. doi:10.1057/s41599-024-03864-y.

Maeda, T.; and Stark, L. 2025. Anthropomorphism as Social Affordance: Charting the Co-Animation of Chatbots into Social "Agents". Proceedings of the AAAI/ACM Conference on AI, Ethics, and Society 8(2): 1661–1673. doi:10.1609/aies.v8i2.36664.

Meadi, M. R.; Sillekens, T.; Metselaar, S.; van Balkom, A.; Bernstein, J.; and Batelaan, N. 2025. Exploring the Ethical Challenges of Conversational AI in Mental Health Care: Scoping Review. JMIR Mental Health 12: e60432. doi:10.2196/60432.

NIST (National Institute of Standards and Technology). 2023. Artificial Intelligence Risk Management Framework (AI RMF 1.0). Gaithersburg, MD: National Institute of Standards and Technology. doi:10.6028/NIST.AI.100-1.

Norhashim, H.; and Hahn, J. 2024. Measuring Human-AI Value Alignment in Large Language Models. Proceedings of the AAAI/ACM Conference on AI, Ethics, and Society 7(1): 1063–1073. doi:10.1609/aies.v7i1.31703.

Ouyang, L.; Wu, J.; Jiang, X.; Almeida, D.; Wainwright, C.; Mishkin, P.; Zhang, C.; Agarwal, S.; Slama, K.; Ray, A.; Schulman, J.; Hilton, J.; Kelton, F.; Miller, L.; Simens, M.; Askell, A.; Welinder, P.; Christiano, P.; Leike, J.; and Lowe, R. 2022. Training Language Models to Follow Instructions with Human Feedback. In Advances in Neural Information Processing Systems 35: 27730–27744.

Peter, O.; and Devlin, K. 2025. Decentralising LLM Alignment: A Case for Context, Pluralism, and Participation. Proceedings of the AAAI/ACM Conference on AI, Ethics, and Society 8(2): 1988–1999. doi:10.1609/aies.v8i2.36690.

Purpura, A.; Wadhwa, S.; Zymet, J.; Gupta, A.; Luo, A.; Rad, M. K.; Shinde, S.; and Sorower, M. S. 2025. Building Safe GenAI Applications: An End-to-End Overview of Red Teaming for Large Language Models. In Proceedings of the 5th Workshop on Trustworthy NLP (TrustNLP 2025), 335–350. Albuquerque, New Mexico: Association for Computational Linguistics. doi:10.18653/v1/2025.trustnlp-main.23.

Rao, A.; Keller, A.; Kalra, N.; Steed, R.; Kwegyir-Aggrey, K.; Klyman, K.; Staheli, D.; and Bergman, A. 2026. Challenges to the Monitoring of Deployed AI Systems: Center for AI Standards and Innovation. NIST AI 800-4. Gaithersburg, MD: National Institute of Standards and Technology. doi:10.6028/NIST.AI.800-4.

Sharma, M.; Tong, M.; Korbak, T.; Duvenaud, D.; Askell, A.; Bowman, S. R.; Durmus, E.; Hatfield-Dodds, Z.; Johnston, S. R.; Kravec, S. M.; Maxwell, T.; McCandlish, S.; Ndousse, K.; Rausch, O.; Schiefer, N.; Yan, D.; Zhang, M.; and Perez, E. 2024. Towards Understanding Sycophancy in Language Models. In The Twelfth International Conference on Learning Representations.

Susser, D.; Roessler, B.; and Nissenbaum, H. 2019. Technology, Autonomy, and Manipulation. Internet Policy Review 8(2): 1–22. doi:10.14763/2019.2.1410.

Sweller, J. 1988. Cognitive Load During Problem Solving: Effects on Learning. Cognitive Science 12(2): 257–285. doi:10.1207/s15516709cog1202_4.

Turner, C.; and Eisikovits, N. 2026. Programmed to Please: The Moral and Epistemic Harms of AI Sycophancy. AI and Ethics 6: 168. doi:10.1007/s43681-026-01007-4.

Vasconcelos, H.; Jörke, M.; Grunde-McLaughlin, M.; Gerstenberg, T.; Bernstein, M. S.; and Krishna, R. 2023. Explanations Can Reduce Overreliance on AI Systems During Decision-Making. Proceedings of the ACM on Human-Computer Interaction 7(CSCW1): Article 129, 1–38. doi:10.1145/3579605.

Waytz, A.; Heafner, J.; and Epley, N. 2014. The Mind in the Machine: Anthropomorphism Increases Trust in an Autonomous Vehicle. Journal of Experimental Social Psychology 52: 113–117. doi:10.1016/j.jesp.2014.01.005.

Weidinger, L.; Uesato, J.; Rauh, M.; Griffin, C.; Huang, P.-S.; Mellor, J.; Glaese, A.; Cheng, M.; Balle, B.; Kasirzadeh, A.; Biles, C.; Brown, S.; Kenton, Z.; Hawkins, W.; Stepleton, T.; Birhane, A.; Hendricks, L. A.; Rimell, L.; Isaac, W.; Haas, J.; Legassick, S.; Irving, G.; and Gabriel, I. 2022. Taxonomy of Risks Posed by Language Models. In Proceedings of the 2022 ACM Conference on Fairness, Accountability, and Transparency, 214–229. New York: ACM. doi:10.1145/3531146.3533088.

Zhang, J.; Zhou, Y.; Liu, Y.; Li, Z.; and Hu, S. 2024. Holistic Automated Red Teaming for Large Language Models through Top-Down Test Case Generation and Multi-Turn Interaction. In Proceedings of the 2024 Conference on Empirical Methods in Natural Language Processing, 13711–13736. Miami, Florida: Association for Computational Linguistics.